\documentclass[letterpaper,english,prd,nofootinbib,showpacs,preprintnumbers,fleqn,floatfix]{revtex4}
\usepackage{array}
\usepackage{graphicx}

\makeatletter

\usepackage{babel}
\makeatother
\begin{document}

\newcommand{\prescr}[1]{{}^{#1}}
\newcommand{\ra}{\rightarrow}
\newcommand{\alphas}{\alpha_{s}}


\title{Internal conversions in Higgs decays to two photons.}

\author{Ana Firan and Ryszard Stroynowski}

\affiliation{Department of Physics, Southern Methodist University, Dallas,
Texas 75275-0175, U.S.A.\\
}

\begin{abstract}
We evaluate the partial widths for internal conversions in the Higgs
decays to two photons. For the Higgs masses of interest at the LHC in the range
of 100-150 GeV, the conversions to pairs of fermions represent a significant
fraction of Higgs decays.

\end{abstract}

\date{\today{}}

\maketitle


The Higgs mechanism~\cite{Higgs} has been introduced into the Standard Model~\cite{Weinberg,Salam,Glashow} to explain electroweak symmetry breaking and the masses of the fundamental particles.
In its simplest form the Standard Model requires a single neutral observable boson H. The search for
the Higgs boson has been one of the main motivations for the construction
of the Large Hadron Collider (LHC).

The theoretical properties of the Standard Model Higgs boson have been extensively studied~\cite{Hhunters}.
Its production mechanisms, coupling and most of the major decays have been well understood.
The mass of the Higgs boson remains the only free
parameter. The lower limit on the mass - $m_{H}>$ 114.4 GeV at 95\% CL - has been established in the direct searches
done by the LEP experiments~\cite{pdg}.
The global fits to the numerous data on electroweak processes show a strong preference for the low mass of the Higgs.
The current best fit value is $m_{H}<$ 186 GeV at 95\%CL~\cite{pdg}.
Low mass Higgs will decay predominantly to a pair of fermions or a pair of bosons. In the LHC experiments such decays
will have to be disentangled from the copious background from QCD processes.
One of the most promising channels for its observation at the LHC is the decay $H\to\gamma\gamma$. In the low Higgs mass range,
this  decay has a relatively small branching fraction but is  also expected to have low background rate. It has been used as
a benchmark in optimization of the ATLAS and CMS detectors and in estimates of the discovery potential~\cite{tdr}.

\begin{figure}[h]
\begin{center}
\includegraphics[width=7cm]{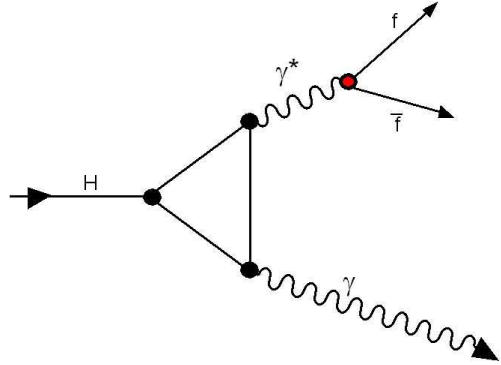}
\end{center}
\caption{The Feynman diagram for the Higgs decay with internal conversion. \label{fig:Feyn}}
\end{figure}

In this note we point out that Higgs decay into two photons may proceed via the internal conversion process analogous to the
Dalitz decay of a neutral pion~\cite{KrollWada, Miyazaki} (see Fig.~\ref{fig:Feyn}$\,$). Here, the internal conversion refers to the decay of a virtual photon, $\gamma^*$,
to a pair of fermions, where the virtual photon mass
can range up to the mass of the Higgs.  In contrast to the case of a neutral pion, the choice of the fermion
type is thus not limited to electrons only but will include all charged leptons and all quarks allowed by the kinematics.
The running effective coupling of the virtual photon to the fermion pair has to be evaluated at the mass of the virtual photon, q~\cite{Barger}:

$${\alpha_{eff}(q^2)} = {{\alpha_0}\over{{1-{{{\alpha_o}\over{3\pi}}\sum_{i}e_{i}^2\Theta(q^2-4m_{i}) ln\left( {q^2}\over 4m_{i}^2\right)}}}},$$

\noindent where $\alpha_0=1/137$ and $m_{i}$ denotes the mass of the fermion in the Callan-Symanzik beta function.

The easiest way to evaluate the rate for such internal conversions is to calculate the ratio of the Higgs decay rate to
a photon and a virtual photon and  decay to a pair of real photons.

$${\rho} = {{\Gamma(H\to\gamma\gamma^*)}\over{{\Gamma(H\to\gamma\gamma)}}},$$
where the $\Gamma$ refers to partial decay width.

In this ratio the terms due to the loop integration cancel out. The value of $\rho$ is given by

$$\rho = {4\over{3\pi}}
\int_{2m_{f}}^{m_{H}} \alpha_{eff} (q^2)
\left( 1- \frac{q^2}{m_{H}^2}  \right)^3
\left( 1- \frac{4m_{f}^2}{q^2}  \right)^{1/2}
\left( 1+ \frac{2m_{f}^2}{q^2} \right)
{dq\over {q}}
$$
where $m_{f}$ is the final state fermion mass.
The corresponding partial width for each channel is
$\Gamma_{i} = \rho\times Br(H\to\gamma\gamma)\times\Gamma_{tot},$
where $\Gamma_{tot}$ denotes the total width.
 The results for three values of the Higgs mass are listed in Table 1 and illustrated in Fig.2.

\begin{table}[htbp]
\begin{center}
\begin{tabular}{|c|c|c|c|c|c|c|} \hline

$ Higgs Mass$ &\multicolumn{2}{|c|}{$m_H$=120GeV}  & \multicolumn{2}{|c|}{$m_H$=150GeV} & \multicolumn{2}{|c|}{$m_H$=180GeV} \\ \hline

$Channel$                &  $\rho$   &  $Branching Fraction$                  &  $\rho$  &   $Branching Fraction$                  &  $\rho$& $Branching Fraction$               \\ \hline
$H\to e^+e^-\gamma$      &  0.0333   &  ${{71.38\times10}^{-6}}$  &  0.0340  &   ${{47.12\times10}^{-6}}$ & 0.0346 & ${{3.4\times10}^{-6}}$  \\
$H\to \mu^+\mu^-\gamma$  &  0.0167   &  ${{35.90\times10}^{-6}}$  &  0.0174  &   ${{24.19\times10}^{-6}}$  & 0.0180 & ${{1.79\times10}^{-6}}$ \\
$H\to \tau^+\tau^-\gamma$&  0.0078   &  ${{16.77\times10}^{-6}}$  &  0.0086  &   ${{11.81\times10}^{-6}}$  & 0.0091 & ${{0.91\times10}^{-6}}$ \\
$H\to u \bar{u}\gamma$   &  0.0211   &  ${{45.36\times10}^{-6}}$  &  0.0220  &   ${{30.58\times10}^{-6}}$  & 0.0229 & ${{2.28\times10}^{-6}}$ \\
$H\to d \bar{d}\gamma$   &  0.0053   &  ${{11.39\times10}^{-6}}$  &  0.0055  &   ${{7.64\times10}^{-6}}$   & 0.0057 & ${{0.57\times10}^{-6}}$ \\
$H\to s \bar{s}\gamma$   &  0.0040   &  ${{8.38\times10}^{-6}}$   &  0.0042  &   ${{5.83\times10}^{-6}}$   & 0.0044 & ${{0.44\times10}^{-6}}$ \\
$H\to c \bar{c}\gamma$   &  0.0123   &  ${{26.44\times10}^{-6}}$  &  0.0132  &   ${{18.35\times10}^{-6}}$  & 0.0140 & ${{1.39\times10}^{-6}}$ \\
$H\to b \bar{b}\gamma$   &  0.0018   &  ${{3.87\times10}^{-6}}$   &  0.0020  &   ${{2.78\times10}^{-6}}$   & 0.0022 & ${{0.22\times10}^{-6}}$ \\
\hline
$ Total$                 &  0.1022   &  ${{219\times10}^{-6}}$    &  0.1070  &   ${{148\times10}^{-6}}$    & 0.1110 & ${{11\times10}^{-6}}$   \\
\hline
\end{tabular}
\caption{Values of the ratio $\rho$ and branching fractions for the Higgs decays to two photons with single internal conversions.
 \label{tab:ht}
}
\end{center}
\end{table}

In this calculation we take into account the color factors for the quarks, the charge dependence of the couplings and assume the lower limit of the
mass integration to be equal to the lowest mass of a physical hadron produced in the decay, i.e., pion mass for
the $u$ and $d$ quarks, kaon mass for the strange quark; we use the particle Data Group values for the masses of $c$
and $b$ quarks~\cite{pdg}.

\begin{figure}[h]
\begin{center}
\includegraphics[width=10cm]{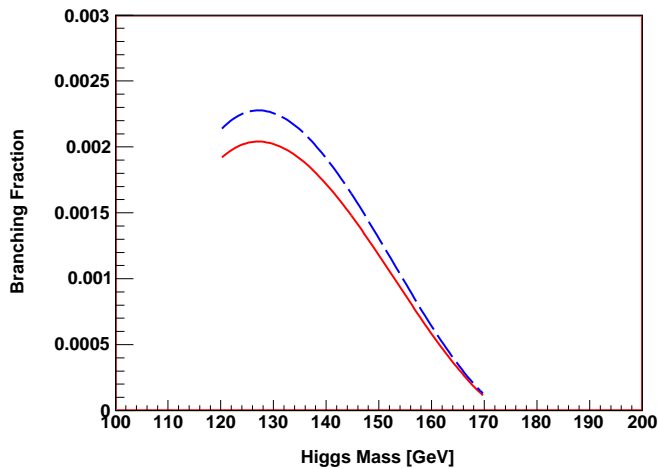}
\end{center}
\caption{The shift in the $Br(H\to\gamma\gamma)$ due to the  Dalitz decay correction.
The dotted line represents the branching fraction without the Dalitz decay correction and the solid line takes into
account the correction. \label{fig:BrFr}}
\end{figure}

As can be seen, in the region of interest to the LHC the total Dalitz decay rate of the neutral Higgs is about 10\% of the
$H\to\gamma\gamma$ branching fraction. Future LHC experiments should include the corresponding correction in their
respective Monte Carlo programs.

Finally, we note that the Higgs Dalitz decay to fermions results in the same final states as for the $H\to Z\gamma$ decay.
The effects of the interference of the corresponding amplitudes have not been yet evaluated.

\section*{Acknowledgments}

We thank F.~Paige and W.J.~Marciano for valuable discussions and helpful comments.
The work was supported by the U.S. Department of
Energy under grant DE-FG03-95ER40908 and the Lightner-Sams Foundation.




\end{document}